\documentclass[aps,prd,twocolumn,showpacs,superscriptaddress,nofootinbib,preprintnumbers]{revtex4-2}

\usepackage[pdftex,colorlinks]{hyperref}

\usepackage{booktabs}

\usepackage[dvipsnames]{xcolor}
\definecolor{phthaloblue}{rgb}{0.0, 0.06, 0.54}

\hypersetup{
    colorlinks=true,
    linkcolor=MidnightBlue,
    citecolor=MidnightBlue,
    filecolor=MidnightBlue,
    urlcolor=MidnightBlue,
    }
\usepackage[pdftex]{graphicx}
\usepackage{bm}
\usepackage{cancel}
\usepackage{here}
\usepackage{comment,braket}

\usepackage{amsmath}
\usepackage{graphics}
\usepackage{amsfonts}
\usepackage{amssymb}
\usepackage{latexsym}
\usepackage[dvipsnames]{xcolor}
\usepackage{natbib}
\usepackage[normalem]{ulem}

\usepackage[utf8]{inputenc}
\DeclareUnicodeCharacter{2212}{-}
\usepackage{orcidlink}
\usepackage{tabularx}

 \newcommand{\brac}[2]{ \left( \frac{#1}{#2} \right) } 

\newcommand{\be}{\begin{eqnarray}}
\newcommand{\ee}{\end{eqnarray}}

\newcommand{\beq}{\begin{equation}}
\newcommand{\eeq}{\end{equation}}

\newcommand{\DR}[1]{{\color{ForestGreen}\textbf{[DR: #1]}}}

\newcolumntype{Y}{>{\centering\arraybackslash}X}

\begin{document}

  \preprint{\tt  FERMILAB-PUB-26-0191-T}

\title{Primordial Neutron Stars}

\author{Gordan Krnjaic\,\orcidlink{0000-0001-7420-9577}}
\email{krnjaicg@fnal.gov}
\affiliation{Theoretical Physics Division, Fermi National Accelerator Laboratory, Batavia, IL USA}
\affiliation{Department of Astronomy \& Astrophysics, University of Chicago, Chicago, IL USA}
\affiliation{Kavli Institute for Cosmological Physics, University of Chicago, Chicago, IL USA}

\author{Duncan Rocha\,
\orcidlink{0000-0002-8263-7982}}
\email{drocha@uchicago.edu}
\affiliation{Kavli Institute for Cosmological Physics, University of Chicago, Chicago, IL USA}
\affiliation{Department of Physics, University of Chicago, Chicago, IL USA}

\author{Huangyu Xiao}
\email{hxiao3@bu.edu}
\affiliation{Theoretical Physics Division, Fermi National Accelerator Laboratory, Batavia, IL USA}

\affiliation{Kavli Institute for Cosmological Physics, University of Chicago, Chicago, IL USA}

\affiliation{
Physics Department, Boston University, Boston, MA, USA
}

\affiliation{Department of Physics, Harvard University, Cambridge, MA, USA
}

\date{\today}

\begin{abstract}
    We propose a novel cosmological scenario in which  baryonic neutron stars could plausibly form in the early universe. 
    If baryogenesis initially produces an excessively-large baryon asymmetry, $Y_B  \gg 10^{-10},$ the baryonic mass inside the horizon can exceed the minimum neutron star mass before big bang nucleosynthesis (BBN). While this large asymmetry is present, non-relativistic baryons can dominate the universe and enhanced density perturbations on small scales can gravitationally  collapse Hubble patches shortly after horizon re-entry. For some initial perturbations, just below the threshold for black hole formation, this collapse will be arrested only by nuclear pressure, possibly resulting in neutron star formation. Afterwards, there must be a large entropy injection to restore the observed baryon asymmetry, $Y_B \sim 10^{-10}$, and preserve the successful predictions of standard BBN. 
Unlike neutron stars that form from stellar collapse, primordial neutron stars can, in principle, be as light as $\sim 0.1 M_\odot$, limited only by the nuclear equation of state.
\end{abstract}

\bigskip
\maketitle

\section{Introduction}
Neutron stars were originally proposed in 1934 by Baade and Zwicky \cite{Baade:1934wuu},  and first observed in 1967 by Bell and Hewish \cite{Hewish:1968bj}. These compact objects are typically of order the solar mass with $\sim$ 10 km-scale radii and consist mainly of degenerate neutrons whose decays are inhibited by Pauli blocking. All known neutron stars are believed to have formed from the end stage of stellar evolution, during which a star with mass above the Chandrasekhar limit of $\sim 1.4 M_\odot$, exhausts its nuclear fuel and undergoes gravitational collapse, arrested only by neutron degeneracy pressure; for a review, see Ref. \cite{Baym:2017whm}.

Although all observed black holes are also believed to have formed from stellar collapse, pioneering work by Novikov and Zel'dovich \cite{Zeldovich:1967lct}, and Carr and Hawking \cite{Carr:1974nx} argued that black holes might also form from the collapse of cosmological perturbations in the high-density early universe. This scenario allows primordial black hole (PBH) formation under a wide range of initial conditions, provided that small-scale density perturbations exceed a critical threshold at early times. For a reviews of PBH formation, see Refs. \cite{Yoo:2022mzl,Green:2020jor}.

In this {\it Letter}, we show that primordial {\it neutron stars} (PNS) may also form in the early universe if the following conditions are satisfied: 1) baryogenesis initially creates an excessively-large baryon asymmetry, 2) $\sim$ 10 km scale density perturbations are greatly enhanced, and 3) an entropy transfer adjusts the baryon asymmetry to its current value in between PNS formation and big bang nucleosynthesis (BBN). 
 As long as the radiation temperature after the entropy injection remains below the quantum chromodynamics (QCD) confinement scale, $\Lambda_{\rm QCD}\sim 200$ MeV, PNS may survive to the present day. 

Our goal is not to demonstrate PNS formation from first principles. Rather, we identify a set of cosmological conditions under which such formation may be possible and argue that the idea is sufficiently plausible to warrant dedicated numerical study. In particular, our analysis does not determine the nonlinear collapse outcome, the precise threshold separating PBH and PNS collapse, or the critical overdensity required to achieve hydrostatic equilibrium.

\section{Basic Ingredients}

\begin{figure*}[t!]    \includegraphics[width=0.95
    \linewidth]{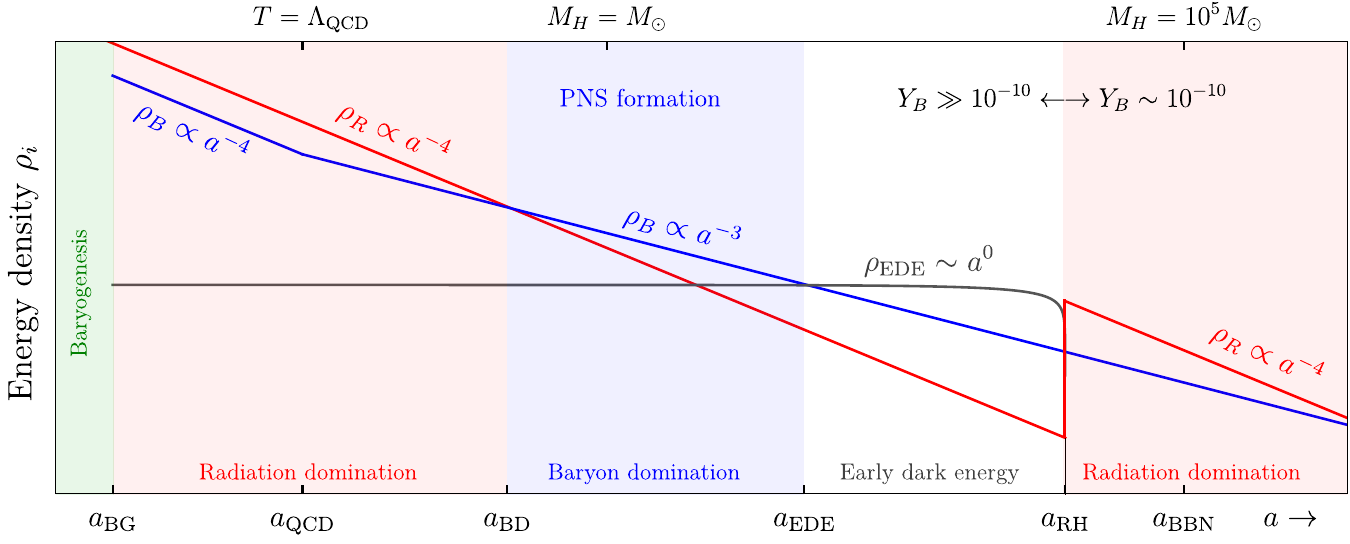}
    \caption{Schematic timeline of an example scenario that allows for primordial neutron star formation. Compared to standard cosmology, this cosmic history invokes two new ingredients: a large baryon yield $Y_B \gg 10^{-10}$ just after baryogenesis and an early dark energy component whose later entropy injection restores radiation domination and yields the observed value $Y_B(a_{\rm RH}) = 8.7 \times 10^{-11}$ before BBN.  At scale factor $a = a_{\rm BG}$ a hot thermal bath with a large baryon asymmetry is created with $Y_B(a_{\rm BG}) \gg 10^{-10}.$ At $a = a_{\rm QCD}$, corresponding to $T = \Lambda_{\rm QCD}$, the baryon number is confined into protons and neutrons whose energy density begins to redshift like non-relativistic matter, eventually dominating the universe at $a = a_{\rm BD}$. During baryon domination, the horizon mass $M_H$ exceeds the minimum value required for neutron star formation and horizon patches with large overdensities collapse to form these objects. Finally, to ensure that $Y_B$ acquires its observed value before BBN, there must be an entropy transfer at $a = a_{\rm RH}$ to significantly decrease $Y_B$.
    }
    \label{fig:timeline}
\end{figure*}

In order for  PNS to form, the baryonic mass inside the horizon must satisfy $ M_B \gtrsim 0.1 M_\odot$ \cite{Bordbar:2006ij,Potekhin:2013qqa,Belvedere:2013hla,Yudin:2022nkm}.
Standard cosmology assumes that the early universe is consistently radiation dominated after inflation with a baryon yield $Y_B \sim 10^{-10}$ in between baryogenesis and BBN. 
 Under these assumptions, the horizon only contains sufficient baryonic mass ($M_B$)
  at late times, when
    \be
    \label{eq:MH}
    M_B = \frac{4\pi m_p s Y_B}{3 H^3}   \approx 0.3 M_\odot
    \brac{Y_B}{10^{-10} }
    \brac{500 \, \rm keV}{T}^3,~~~
    \ee
   where $m_p$ is the proton mass, $s$ is the entropy density, and $H$ is the Hubble rate. Thus, causality requires that we wait until after BBN before any PNS could ever hope to form. 
    
    However, structure formation after BBN is well understood: the first baryonic structures form around redshift $z \sim 20$ and neutron stars only form through stellar evolution  \cite{Barkana:2000fd}. In principle, it might also be possible to form neutron stars from collapsing primordial density perturbations at these late times, but realizing the order-unity overdensities required for collapse  on cosmological scales conflicts with cosmic microwave background (CMB) and large scale structure (LSS) bounds on the primordial power spectrum \cite{Bringmann:2025cht}.

Fortunately, large primordial perturbations can viably be realized on much smaller scales in the very early universe, and these are commonly invoked in PBH formation mechanisms \cite{Green:2020jor}. However, from Eq. \eqref{eq:MH}, the horizon does not contain sufficient baryonic mass at early times  unless $Y_B$ can temporarily be increased well above its measured value before BBN. Thus, PNS formation requires the following ingredients: 

\begin{itemize}
    \item {\bf Enhanced Density Perturbations:} On horizon scales that enclose at least $\sim  0.1 M_{\odot}$ of mass, cosmological perturbations must be greatly enhanced relative to the nominal $\delta \rho/\rho \sim 10^{-5}$ size, which is observed on large scales. This requirement is common to both PNS and PBH formation from the collapse of large super-horizon fluctuations.
    
    \item {\bf Large Baryon Number:} Baryogenesis must initially create a baryon-to-photon ratio much larger than the observed value at BBN, $Y_B \gg 10^{-10}$. This requirement ensures that the horizon contains sufficient baryonic mass to form a neutron star before BBN; from Eq.~\eqref{eq:MH}, we can see that in standard cosmology this first occurs well after BBN. For sufficiently large baryon number at temperatures $T \lesssim \Lambda_{\rm QCD}$, the universe quickly becomes dominated by non-relativistic baryons.

    \item {\bf Entropy Injection:} Since we demand an initially-large value of $Y_B$ to increase the baryon density before BBN, we also need to restore the observed value of $Y_B \sim 10^{-10}$ after PNS might form, but before BBN begins. Thus, we require  an entropy injection from some additional fluid to reduce the baryon-to-photon ratio without creating or destroying baryons.

\end{itemize}

Realizing these ingredients necessarily involves a baryogenesis mechanism that yields $Y_B \gg 10^{-10}$ and an additional cosmic fluid that injects entropy into the plasma at $t \lesssim 1$ s,  so that $Y_B \sim 10^{-10}$ in time to preserve the successful predictions of standard BBN.

\section{Cosmological Model}
In this section we realize the ingredients outlined above in a concrete and viable cosmological model that follows the sequence of events 
depicted in Fig. \ref{fig:timeline}. The key non-standard ingredients are a large baryon asymmetry and an additional component of the universe that redshifts like dark energy and is later depleted to produce entropy before BBN.

\subsubsection{Primordial Fluctuations}

Before the timeline in  Fig. \ref{fig:timeline} begins, we assume 
that the primordial fluctuations generated during inflation are enhanced 
 on small scales corresponding to $M_H \sim M_\odot$.  
  There are many viable mechanisms that can yield such enhancements (see Ref. \cite{Ragavendra:2023ret} for a review), and any one of these suffices as long as large-scale modes with $k \gtrsim 10^{-3}$ Mpc$^{-1}$ are unaffected.
  

\subsubsection{Baryogenesis}
To ensure that there is a sufficient baryon energy density inside the causal horizon before BBN, the baryon yield ratio produced during baryogenesis must initially be much larger than the present-day measured value $Y_B = (8.7 \pm 0.04)  \times 10^{-11}$ \cite{ParticleDataGroup:2022pth,Planck:2018vyg}. For our purposes, the details of baryogenesis are not important as long as a large asymmetry can be produced. However, most known mechanisms suffer from significant washout effects, which limit the baryon density that can be realized \cite{Bodeker:2020ghk,Riotto:1998bt}.

Fortunately, Affleck-Dine baryogenesis offers an important counterexample to this general behavior, thereby allowing a large asymmetry to arise naturally \cite{Affleck:1984fy}. In this scenario, a complex scalar field $\phi = \frac{r}{\sqrt{2}} e^{i\theta}$  with baryon number is displaced from its minimum after inflation. If the potential $V(r,\theta)$ has both baryon-preserving and baryon-violating terms, $\phi$ dynamically evolves to acquire a large baryon number, which is transferred to SM fields upon decay. For an appropriate choice of potential,  Affleck-Dine can naturally accommodate a large baryon number \cite{Rubakov:2017xzr}, which yields $Y_B(a_{\rm BG}) \gg 10^{-10}$ when baryogenesis ends. 

Concretely, if $\phi$ dominates the universe and decays through baryon-preserving interactions, it creates a thermal bath of initial temperature $T > \Lambda_{\rm QCD}$ at $a = a_{\rm BG}$  (see Fig. \ref{fig:timeline}), and transfers the baryon asymmetry to SM fields. Assuming instantaneous $\phi \to$ SM decays, the net baryon number density is 
\be
\label{eq:n_B}
n_{B}  \approx \frac{1}{H} \frac{\partial V}{\partial \theta} = \, 
\frac{N_F}{3 \pi^2} \mu ( \pi^2 T^2 + \mu^2) ,~~~~
\ee
where $H$ is the Hubble rate, $N_F$ is the number of light quark flavors at temperature $T$,
$\mu$ is the baryon chemical potential,\footnote{If $T_{\rm BG} \gtrsim 100$ GeV, electroweak symmetry is restored and the baryon asymmetry must instead be a $B-L$ asymmetry to avoid sphaleron washout, but the qualitative argument is unchanged.}, 
the total energy density is 
\be
\label{eq:rho_phi}
\rho_\phi = \frac{ g_\star \pi^2}{30}T^4  + \begin{cases}
\frac{3 N_F }{ 4 } \mu^2 T^2 &(\mu \ll T)\\  
\frac{3 N_F}{8 \pi^2}\mu^4  &(\mu \gg T)\\ 
\end{cases}~,~~~
\ee
where $g_\star$ is the effective number of relativistic species, and
the entropy density is 
\be
\label{eq:entropy}
s = \frac{2\pi^2 g_{\star, S} }{45} T^3 + \frac{7N_F}{24} \mu^2 T,
\ee
where $g_{\star, S}$ is the number of effective relativistic species in entropy.  
For given values of $\rho_\phi$ and $n_B$, which depend on $V$ and the initial value of  $\phi$ at the time of decay,  Eqs. \eqref{eq:n_B} and \eqref{eq:entropy} can be solved simultaneously to obtain $Y_B(a_{\rm BG})= n_B/s \gg 10^{-10}$. 

\subsubsection{Radiation Domination}
Once the baryon asymmetry is transferred to SM fields, the universe is dominated by radiation with a large baryon number. As the radiation cools, $T, \mu\propto a^{-1}$ and strongly interacting matter enters the hadronic  phase for $\sqrt{T^2 + (\mu/7)^2} \lesssim \Lambda_{\rm QCD}$, which approximates the QCD phase boundary \cite{Kronfeld:2012uk}. Inside this region, quarks and gluons become confined and baryon number is stored in non-relativistic protons and neutrons, so $\bar \rho \propto a^{-3}$. Note that, due to the large asymmetry during this epoch, the  time-temperature relation is $\mu$-dependent
 \be
H = \frac{1}{2t} = \sqrt{\frac{8\pi G}{3} \!
\left(
 \frac{g_\star\pi^2}{30}T^4 + 
\frac{3N_F}{8 \pi^2}\mu^4  \right)}~~,
\ee
where we have used the $\mu \gg T$ limit from Eq.~\eqref{eq:rho_phi}; in the opposite regime, the usual $t \sim 1/T^2$ relation holds. 

\subsubsection{Baryonic Matter Domination}

Once confinement occurs at $a = a_{\rm QCD}$, quarks and gluons are confined into baryons, which redshift like non-relativistic matter. If $Y_B(a_{\rm BG}) \gtrsim 0.2$, then baryons dominate the universe immediately upon confinement, so $a_{\rm QCD} = a_{\rm BD}$, but if $Y_{B}(a_{\rm BD}) \lesssim 0.2$, then $a_{\rm BD} > a_{\rm QCD}$ and depends on $Y_B(a_{\rm BG}).$

During this era, the energy density is
\be
\rho = \rho_{B} + \rho_R + \rho_{\rm EDE}~~,
\ee
where $\rho_{B} =  m_p Y_B s  \propto a^{-3}$
is now the non-relativistic baryon energy density, $\rho_R$ is the subdominant radiation density,
and
$\rho_{\rm EDE} \propto a^0$ is an initially-subdominant early dark energy (EDE) density.\footnote{In Refs. \cite{Karwal:2016vyq,Poulin:2018cxd} a subdominant EDE substance is invoked to modify the expansion rate near matter-radiation equality to address the Hubble tension. By contrast, here the EDE must be active before BBN and also eventually dominate the universe before transferring its entropy to the SM.} At the beginning of  this era, the radiation and EDE fractions satisfy
\be
f_{\rm EDE} \equiv \frac{ \rho_{\rm EDE}(a_{\rm BD}) }{  \rho_{B}(a_{\rm BD} ) } \ll 1~,~
\ee
and the value of $a_{\rm BD}$ depends on $\mu$ since, for fixed $\rho_\phi$ in Eq.~\eqref{eq:rho_phi}, increasing $\mu$ results in earlier baryon domination. Note that, by charge neutrality, there is a compensating charged lepton asymmetry, but its energy density is a small correction to $\bar \rho$ and is negligible.

During this period of matter domination, enhanced density perturbations on $k\sim 10^4 \rm \, Mpc^{-1}$ comoving scales re-enter the horizon when it encloses $M_H \gtrsim M_\odot$ of baryonic mass. As discussed below in Sec. \ref{sec:PNS_formation}, for sufficiently large perturbations, PNS may result from the collapse of these horizon patches. 

\subsubsection{Early Dark Energy Domination}

\begin{figure}[t]
\hspace{-0.4 cm}
\includegraphics[width=1.03
    \linewidth]{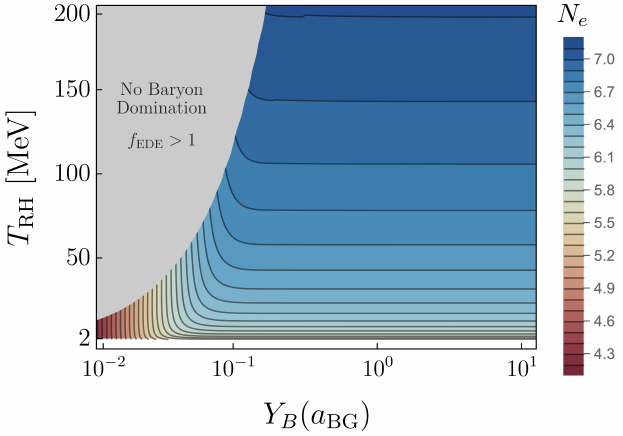}
    \caption{Parameter space for our inputs:  $T_{\rm RH}$, the radiation temperature after EDE domination ends, and $Y_B(a_{\rm BG})$, the initial baryon yield just after baryogenesis. The contours represent the number of e-folds of cosmic expansion during EDE domination, $N_e$, needed to restore the asymmetry to the observed value $Y_B(a_{\rm RH}) = 8.7 \times 10^{-11}$ after the EDE era. Note that for $Y_B(a_{\rm BG}) \lesssim 10^{-2}$, it is not possible to achieve baryon domination for a viable choice of $T_{\rm RH}$. For $Y_B(a_{\rm BG}) \gtrsim 10^{-1}$, baryon domination is nearly instantaneous once the universe cools to $T \sim \Lambda_{\rm QCD}$. In the gray shaded region, there is no baryon domination $f_{\rm EDE}$. From Eq.~\eqref{eq:YB_RH}, at large $Y_B \gg 0.2$, the asymmetry depends only on $T_{\rm RH}$ and $N_e$, so the contours are constant in $Y_B$; see appendix \ref{app:yield-calc} for a discussion.}
    \label{fig:param-space}
\end{figure}

Since $\rho_{\rm EDE} / \rho_{B} \propto a^3$, the epoch of baryon domination eventually ends as the EDE comes to dominate the early universe at scale factor 
$a_{\rm EDE} =  a_{\rm BD}/f^{1/3}_{\rm EDE}$,  
after which the non-relativistic baryons constitute  only a small fraction of the horizon mass and it becomes increasingly difficult to form PNS. 
During EDE domination, the scale factor grows as 
\be
\label{eq:aEDE}
a(t) = a_{\rm EDE} \exp \left[ {H_{\rm EDE} (t- t_{\rm EDE}) }\right],
\ee
and the Hubble rate  satisfies 
$ H^2_{\rm EDE} \approx  8\pi G \rho_{\rm EDE}/3$,
which is approximately constant until $\rho_{\rm EDE}$ is converted to radiation and reheats the universe.

In this scenario, the EDE fluid can be identified with the potential energy of an additional scalar singlet field $S$, which is initially trapped in a false vacuum at $S = 0$, so $\rho_{\rm EDE} = V(0)$.  This era ends at $a= a_{\rm RH}$ when $S$ tunnels to the true minimum at $S = v_S$ and reheats the universe before BBN. Such a late-time phase transitions produce stochastic gravitational wave backgrounds \cite{Christensen:2018iqi,Caprini:2018mtu} with consequences for CMB anisotropies \cite{Greene:2024xgq,Greene:2026one,Zebrowski:2026pye}, but for our purposes it suffices mainly to inject entropy.

\subsubsection{Reheating}

When the $S$ tunnels to its true minimum at $a = a_{\rm RH}$, the subsequent SM radiation density can be written 
\be
\rho_{\rm EDE} \approx \frac{\pi^2 g_\star( T_{\rm RH}) }{30} T_{\rm RH}^4,
\ee
where $T_{\rm RH}$ is the reheat temperature, which must exceed a few MeV to allow for a viable epoch of BBN. By this point in cosmic history, the baryon number density is
\be
n_{B}(a_{\rm RH}) = \frac{\rho_{B}(a_{\rm BD})}{m_p} \brac{a_{\rm BD}}{a_{\rm RH}}^3,
\ee
which is exponentially sensitive to the duration of EDE since $a_{\rm RH} \approx a_{\rm EDE} \exp(N_e)$ from Eq.~\eqref{eq:aEDE}, where $N_e$, is the number of $e$-folds during EDE.

After reheating, a relativistic bath of SM particles is created with reheat temperature that satisfies  $T_{\rm RH} < \Lambda_{\rm QCD}$ to ensure that baryons inside the PNS are not converted into relativistic quarks, which would destroy any PNS formed during baryon domination; we also demand that $T_{\rm RH} \gtrsim$ MeV to allow for standard BBN to proceed \cite{Hasegawa:2019jsa}. If no net baryons are created during reheating, the baryon-to-entropy ratio satisfies
\be
\label{eq:YB_RH}
~~Y_B(a_{\rm RH}) = 
 \frac{n_{B}(a_{\rm RH}) }{ s( a_{\rm RH}) } =
 \frac{3 T_{\rm RH}}{2 m_p}\, e^{-3N_e}
 ~~,~~ 
\ee 
where 
$s(a_{\rm RH})$ is the entropy density at reheating and we have neglected residual radiation in the last step, which is valid for large $Y_B(a_{\rm BG})$ -- see Appendix \ref{app:yield-calc} for a discussion. In Fig. \ref{fig:param-space} we show the parameter space that realizes the observed value $Y_B= 8.7 \times 10^{-11}$ for  viable choices of input parameters.

\begin{figure*}[t!]
    \centering
    \includegraphics[width=0.9\linewidth]{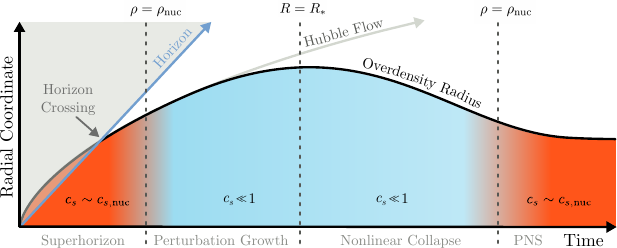}
    \caption{
    Schematic cartoon of PNS formation that tracks the evolution of the overdensity radius from horizon crossing to  collapse in our scenario. The red/blue shaded regions correspond to times at which the baryon density is above/below nuclear density, respectively. The perturbation initially enters the horizon when the sound speed is large, and initially tracks the Hubble flow $R(t) \propto a(t) \propto t^{2/3}$. It continues expanding until the sound speed drops, at which point the perturbation grows, eventually reverses its expansion $(R = R_\star)$, and collapses to a proto-neutron star.
    }
    \label{fig:CollapseCartoon}
\end{figure*}

\section{Heuristic Formation Criteria}
\label{sec:PNS_formation}

 In baryon domination, the horizon mass  is 
 \be
 M_H =  \brac{3}{32 \pi G^3 \rho_B}^{1/2}   \approx 1M_\odot \brac{10^2 \rho_{\rm nuc}}{\rho_B}^{1/2} \!\!\!,~~~
 \ee
 where $\rho_{\rm nuc} \approx 2 \times 10^{14} \rm \, g \, cm^{-3}$, is the nuclear density. When $M_H \in (0.1-2) M_\odot$, 
there is sufficient baryonic mass inside the horizon to potentially form a stable neutron star  \cite{Douchin:2001sv,Bordbar:2006ij}.
 During this matter-dominated era, when density
 fluctuations  
 $\delta(t_0) \equiv \delta_H$
enter the horizon at time $t = t_0$, they initially grow as $\delta \propto a$ and reach a maximum radius $R_*$ at $t= t_*$. At this point, if the interior density satisfies $\rho_B < \rho_{\rm nuc}$, the sound speed can be modeled as

\be
\label{eq:sound-speed-star}
c_{s}^2(t_{*}) \approx \frac{5T_*}{3m_p} + \frac{4\rho_{R*}}{9 \rho_{B*}}  + \frac{1}{3} \left( \frac{ 3\pi^2 \rho_{B*} }{ 16 m_p^4}  \right)^{1/3}\!\!,~~ 
\ee 
where $*$ denotes a quantity evaluated at $t = t_*$, $\rho_R/\rho_B$ is a free parameter set by the large value of $Y_B$ at this time and the last term accounts for electron degeneracy pressure.\footnote{The electron degeneracy term assumes charged lepton fraction $n_e/n_B = 1/2$, but this could possibly be smaller in the $\mu > T$ regime \cite{Moustakidis:2008hg}.}

\begin{table*}[t]
\centering
\begin{tabular}{|l|c c c|c c c c|c c c|}
\cline{2-11}
\multicolumn{1}{c|}{} 
& \multicolumn{3}{c|}{\textbf{Cosmology}} 
& \multicolumn{4}{c|}{\textbf{Horizon Crossing}} 
& \multicolumn{3}{c|}{\textbf{Collapse}} \\
\cline{2-11}
\multicolumn{1}{c|}{\textbf{}} 
& $Y_B$ 
& $T_{\rm RH}$ 
& $N_e$
& $R_H$ 
& $\delta_H$ 
& $T_H$ 
& \!\!\! $M_H$ 
& $R_{*}$ 
& $c^2_{s}(t_*)$ 
& $\tau_{*}$ \\
\hline

\textbf{Benchmark 1}~~~~ 
& 0.5 ~
& 10 \rm MeV~~
& 4.5~
& ~~~3.2 km ~~
& ~~0.06 ~~
&~~ 179\,MeV~~ 
& ~~$1.3 \,M_\odot$ ~~
& $64$\,km ~~
& 0.02 ~~
& $0.003$ \rm s \\

\textbf{Benchmark 2}~~~~ 
& 0.2 ~
& 50 \rm MeV~~
& 6.5~
& ~~~5.0\,km ~~
& ~~0.01 ~~
&~~ 192\,MeV~~ 
& ~~$1.9 \,M_\odot$ ~~
& $532$\,km ~~
& 0.005 ~~
& $0.05$ \rm s \\

\textbf{Benchmark 3}~~~~ 
& 1.3~
& 30 \rm MeV~~
& 5~
& ~~~2.2\,km ~~
& ~~0.15 ~~
&~~ 155\,MeV~~ 
& ~~$0.9 \,M_\odot$ ~~
& $20$\,km ~~
& 0.05 ~~
& $0.0005$ \rm s \\

\hline
\end{tabular}
\caption{
Example model parameters which satisfy the necessary (but not necessarily sufficient) criteria for PNS formation.
Here $T_H$ is the temperature of the universe when the perturbation enters the horizon, which we also take to the be the confinement temperature without loss of essential generality. Since $Y_B$ is large at early times, these benchmarks all predict that baryon domination occurs suddenly at confinement.
Also, the collapse timescale at $R_*$ is  $\tau_{*} = (G \rho_*)^{-1/2}$, where $\rho_*$ is the density of the perturbed region when it reaches its maximum radius.
We assume equal proton--neutron number densities, which affects the electron degeneracy pressure contribution to the sound speed at maximum radius.
See Appendix~\ref{app:formation} for details.
}
\label{tab:PNSExample}
\end{table*}

If the maximum radius exceeds the Jeans scale, $R_* > R_J \sim c_{s,*}/\sqrt{G\rho_{B*}
}$, where $\rho_{B*}$ is the mass density enclosed 
within the perturbation at $t = t_*$, the overdensity will collapse inward. As shown in Appendix \ref{app:formation}, the Jeans criterion is equivalent to the requirement  $\delta_H > c^2_{s,*}$, 
and if the sound speed were to remain constant throughout the infall, 
any perturbation that satisfies this inequality would collapse to form a PBH \cite{Carr:1974nx,Yoo:2022mzl}.

However, as the core density increases to $ \rho_B \sim \rho_{\rm nuc}$,
Eq.~\eqref{eq:sound-speed-star} should be replaced by the nuclear EOS, and 
 the sound speed grows to $c_{s,\rm nuc}^2 \approx 0.2-0.6$ \cite{Fujimoto:2024cyv,Fukushima:2025ujk}. Thus, an overdensity which satisfies $\delta_H > c_s^2(t_*)$ is no longer guaranteed to form a PBH. Rather, for perturbation values in the range 
\be
\label{eq:delta_range}
c_{s}^2(t_*) \lesssim \delta_H \lesssim c_{s, \rm nuc}^2~,
\ee
the overdensity begins to collapse, but may eventually be arrested by a pressure response and form a PNS; however, if $\delta_H > c^2_{s,\rm nuc}$ the nuclear response is insufficient to avert PBH formation. While the precise threshold for PNS formation requires a dedicated simulation, this critical value is expected to lie within the range of Eq.~\eqref{eq:delta_range} and the steps outlined here are schematically depicted in  Fig. \ref{fig:CollapseCartoon}.

In Table \ref{tab:PNSExample} we present concrete benchmark values of model parameters that allow for overdensities to begin their collapse at $t_*$, but whose 
infall could be significantly affected by a pressure wave when the nuclear equation of state stiffens. 
The values in this table can be computed following the prescription in Appendix \ref{app:formation}, which adapts the 
formalism of PBH formation to our scenario and derives the collapse criterion within the assumptions of our scenario. 

The window in Eq.~\eqref{eq:delta_range} should be understood as a necessary but insufficient condition for PNS formation. While collapse initiates for $\delta_H > c_{s}^2(t_*)$ and the stiffening nuclear equation of state at $\rho_B \sim \rho_{\rm nuc}$ provides a physical mechanism for arresting infall, whether a given collapsing patch settles into hydrostatic equilibrium — rather than forming a black hole or ejecting its envelope — depends on nonlinear dynamics that cannot be resolved by this analysis. Additional effects, including heating, neutrino emission, shock formation, turbulence, and matter ejection, are important corrections to our analysis, especially in the case of large maximum radius, $R_* \gg 10$\,km. We therefore present Eq.~\eqref{eq:delta_range} as a plausibility argument motivating dedicated numerical simulation, analogous to early analytic estimates in the PBH literature that preceded modern threshold calculations. 

Furthermore, in the study of PBH formation, the critical collapse threshold should depend sensitively on the shape of the perturbation profile and on the sound speed during formation \cite{Harada:2013epa}. Relative to this scenario, we introduce an  additional complication: the equation of state evolves significantly during infall, from a baryonic fluid with subdominant radiation at horizon crossing to nuclear matter at high density. Whether a given Hubble patch crosses the PBH threshold before or after the equation of state stiffens depends on details of the infall trajectory, which require a dedicated simulation to resolve. We therefore cannot reliably determine what fraction of collapsing patches in the window of Eq.~\eqref{eq:delta_range} form PNS versus PBH.

\section{Surviving the Plasma}
In our scenario, after PNS formation occurs in baryon domination, before an entropy transfer eventually restores radiation domination. 
Since viable BBN requires $T_{\rm RH} > 2$ MeV \cite{Hasegawa:2019jsa}, naively it would seem that a neutron star immersed in such an environment would be destroyed. However, the escape velocity at the surface of a neutron star is $v_{\rm esc} \sim 0.6$, so a typical nucleon needs $\sim 200$ MeV of kinetic energy in order to become unbound. As long as $T_{\rm RH} \lesssim 200$ MeV, individual collisions with the plasma are insufficient to eject nucleons from the object.

Furthermore, at reheating, the energy flux of particles in the plasma is of order $\Phi \sim T_{\rm RH}^3$, so assuming their energy is uniformly absorbed by the PNS, the total energy deposited in a Hubble time is $E_{\rm dep}\sim 4\pi \Phi R^2 T_{\rm RH} H^{-1} \sim 10^{50}$ erg. The binding energy of a solar-mass PNS with radius $R\sim 10$ km, is $E_{\rm bind} \sim GM^2/R \sim 10^{53}$ erg, so the object is expected to avoid disruption upon reheating.

Intriguingly, we note that PBH and PNS are the only SM compact objects which could survive the minimum reheat temperatures required for BBN. All others, including stars and white dwarfs,
have much lower escape velocities and would be destroyed if present at reheating with $T_{\rm RH }\sim$ MeV.

\section{Observational Consequences}
Primordial neutrons stars exhibit several key differences with respect to their astrophysical counterparts:
\begin{enumerate}
    \item {\bf Mass Scale:} The smallest neutron star ever observed was found in the J0453+1559 binary system with mass $(1.174 \pm 0.004) M_\odot$ \cite{Martinez:2015mya}, which is consistent with the theoretical minimum mass, assuming a   core collapse origin \cite{2012ApJ...757...91B}. However, based only on stability criteria, the smallest possible neutron star mass is $\approx 0.1 M_\odot$ \cite{Bordbar:2006ij,Potekhin:2013qqa,Belvedere:2013hla,Yudin:2022nkm}. Since our scenario does not rely on core collapse, it may be possible to form sub-solar PNS given appropriate initial conditions. However, as noted above, a dedicated simulation is necessary to properly determine the formation criteria for this mechanism.

    \item {\bf Temperature \& Magnetic Field:} Since PNS maintain kinetic equilibrium with the CMB for much of cosmic history, but rarely interact with cosmological structures, they are expected to be much colder than their astrophysical counterparts. This makes direct observation very challenging. Similarly, since they are diffuse and rarely encounter other objects to form binaries, PNS are not expected to become pulsars or host appreciable magnetic fields, except in very rare circumstances. Nevertheless, PNS can viably be very abundant compared to other neutron stars, so it may be possible to observe a sub-solar PNS.
    
    \item {\bf Cosmological Abundance:} Since PNS are diffuse and rarely interact with their surroundings, they are expected to have the same spatial distribution as dark matter (DM). As with PBH with $\sim M_\odot$ mass, PNS cannot constitute all of the dark matter due to lensing constraints for solar-mass sized objects \cite{Bringmann:2025cht,Green:2020jor}, but they can be an appreciable sub-component with $\rho_{\rm PNS}/\rho_{\rm DM} \lesssim 10^{-1}$. Note that this estimate assumes that limits on the PBH abundance near $\sim M_\odot$ also apply to PNS; a dedicated study is needed to understand the proper constraints on this scenario.  Nevertheless, even for PNS abundances near this conservative
    upper limit, most of the neutron stars in the galaxy would be primordial.

\end{enumerate}

\section{Discussion}
In this {\it Letter}, we have proposed a cosmological mechanism to plausibly form neutron stars prior to BBN, which to our knowledge has not been previously proposed. We find that that these objects might viably form if the early universe contains large primordial density fluctuations on small scales, a temporarily-large baryon asymmetry, and an entropy injetion to restore the observed asymmetry before BBN begins. However, these are not necessarily unique conditions and other, more minimal scenarios may also be possible. 

While we have identified the necessary conditions for PNS formation from collapsing density perturbations,  there is considerable uncertainty about the details of the collapse process. A dedicated simulation is necessary to determine the precise initial conditions compatible with this scenario, and we leave this to future work. 

We also note that our scenario is compatible with a wide range of dark matter candidates. Production can occur at any point in the timeline before $z \sim 6 \times 10^5$, after which dark matter must be present as a non-relativistic species to account for the observed matter power spectrum \cite{Das:2020nwc}. If dark matter is present during baryon domination, then the sound speed will be decreased, so the numerical values that enter into Eq.~\eqref{eq:delta_range} will change, but the qualitative physics is unchanged. 

Finally, throughout this work, we have assumed that the usual PBH abundance limits near $\sim M_\odot$ apply to our scenario. However, many of these constraints (e.g. gravitational waves from mergers) assume that the signal arises specifically from PBH, so the true limit on the PNS abundance is expected to be weaker. If the true limits are considerably weaker than those for PBH, it might be possible for PNS to constitute most or all of the dark matter abundance, but a dedicated study is necessary to settle this issue.

\bigskip 

\noindent {\bf Acknowledgments}: We thank Torsten Bringmann, Djuna Croon, Josh Foster, Dan Hackett, Aurora Ireland, Gonzalo Hererra, Danimal Hooper, Hank Lamm, Shirley Li,  Yuxin Liu, Andrew Long, Julian Mu\~noz, Maxim Pospelov, Nirmal Raj, Erwin Tanin, Tanner Trickle, Sarunas Verner, and Mike Wagman for helpful conversations.  This document was prepared using the resources of
the Fermi National Accelerator Laboratory (Fermilab),
a U.S. Department of Energy, Office of Science, Office
of High Energy Physics HEP User Facility. Fermilab
is managed by Fermi Forward Discovery Group.  HX is supported by the U.S. Department of Energy under grant
DE-SC0026297.
The authors would like to express a special thanks to the Mainz Institute
for Theoretical Physics (MITP) of the DFG Cluster of Excellence
PRISMA+ (Project ID 39083149), for its hospitality and support.

\appendix

\section{Baryogenesis}
\label{sec:AD-appendix}
In this appendix we present a concrete realization of Affleck-Dine baryogenesis, which can generate a large baryon asymmetry in the early universe \cite{Affleck:1984fy}. 
Following the conventions in Refs.   \cite{Rubakov:2017xzr,Ireland:2024wwn}, we consider a complex field $\phi$ with baryon number $B_\phi$ whose potential is 
\be
V(\phi) = m_\phi^2 |\phi|^2 + \frac{\lambda}{2} |\phi|^4  + \frac{\lambda^\prime}{4}(\phi^4 + \phi^{*4}) ,
\ee
which contains baryon-preserving interactions and baryon-violating interactions proportional to $\lambda^\prime$. In polar coordinates $\phi = \frac{1}{\sqrt{2}} r e^{i \vartheta}$ and the potential is 
\be
V(r,\theta) = \frac{ m_\phi^2}{2} r^2 + \frac{ \lambda}{4} r^4 + \frac{\lambda^\prime}{8} r^4 \cos (4 \vartheta),
\ee
and the explicit baryon violation appears as $\theta$ dependence in the potential.
The baryon number density in $\phi$ can be written $n_B = B_\phi r^2 \dot \vartheta$, which satisfies   
\be
\frac{1}{a^3} \partial_t \left( a^3 n_B\right) =  B_\phi \frac{\partial V}{\partial \vartheta} = \frac{B_\phi \lambda^\prime}{2} \sin(4 \vartheta),
\ee
where $B_\phi$ is the baryon number of $\phi$ and we have used the $\vartheta$ equation of motion. Given initial conditions for $r_i, \vartheta_i$ at some initial time $t_i$, this expression can be integrated to obtain the baryon number stored in the field 
\be
n_B(t) = \frac{B_\phi \lambda^\prime}{2 a^3(t)} \int_{t_i}^t dt^\prime a^3(t^\prime) r(t^\prime)^4 \sin[4\vartheta(t^\prime)]~,
\ee
and we assume that there exists some baryon number preserving $\phi$ interaction to transfer the asymmetry to SM fields and inject entropy into the thermal bath.
In the limit of $\lambda \gg \lambda^\prime$, the total energy in the field is
\be
\rho_\phi  = \frac{1}{2}(\dot r^2 + r^2 \dot \vartheta^2) + \frac{ m^2}{2} r^2 + \frac{ \lambda}{4} r^4 ~,
\ee
and, for simplicity, we assume that 
 $\phi$ is initially displaced from its origin after inflation and dominates the universe
 before decaying to create the SM thermal bath at early times.

Assuming instantaneous $\phi$ decays when $\tau_\phi \sim H^{-1}$, where $\tau_\phi$ is the lifetime (which we regard as a free parameter), the SM radiation bath satisfies the consistency conditions 
\be
n_{B}(\tau_\phi) &=& 
\frac{N_F}{3 \pi^2} \mu ( \pi^2 T^2 + \mu^2) , \\ 
\rho_\phi(\tau_\phi) &=& \frac{ g_\star \pi^2}{30}T^4  +
\frac{3 N_F}{8 \pi^2}\mu^4 ,
\ee
where we have taken the highly-degenerate $\mu \gg T$ limit. To demonstrate that a large baryon asymmetry can be generated in this model, we approximate 
\be
n_B(\tau_\phi) \sim \frac{B_\phi \lambda^\prime}{H}r_i^4 \vartheta_i ~~,~~\rho_\phi(\tau_\phi)  \sim m^2 r_i^2~,
\ee
where $H \sim \sqrt{\rho_\phi}/M_{\rm Pl}$
and solve the system to obtain $\mu$ and $T$ after decay. For $m = 10^5$ GeV, $r_i = 10^3$ GeV, and $\lambda^\prime =10^{-11}$, we find $\mu \approx 10^4$ GeV, $T \approx 10^3$ GeV, resulting in a baryon asymmetry of order $Y_B \sim \mu/T \sim 10$, which can further be dialed to higher values by increasing $\lambda^\prime$. Thus, there is no impediment to obtaining the large asymmetries that are needed for baryon domination at early times.

\section{Early Dark Energy}
Here we present a concrete example of a model 
which satisfies the requirements for the EDE fluid which reheats the universe after baryon domination. We introduce a new SM singlet $S$ which mixes with the Higgs field with the zero-temperature potential \cite{Batell:2012mj}
\be
\label{eq:pot}
V(H,S) &=& - \mu_H^2 |H|^2 + \lambda_H |H|^4  +  \frac{ \mu_S^2}{2}  S^2 - \frac{A}{3} S^3   \nonumber \\
&& +  \frac{ \lambda_S }{4}S^4 + \left( \frac{\lambda_{HS}}{2} S^2  + \kappa S 
\right) |H|^2 + V_0,~~~~~~
\ee
where $H$ is the SM Higgs doublet and $V_0$ is chosen to absorb the vacuum energy in the true vacuum.
After electroweak symmetry breaking $H = [0,  \frac{1}{\sqrt{2}}(v+h)]$, where $h$ is the Higgs boson field and $v = 246$ GeV, there is an $h$-$S$ mixing governed by the parameter
\be
\theta \simeq \frac{v(\kappa + \lambda_{HS} v_S)}{m_h^2 - m_S^2}~,
\ee
through which $S$ inherits matter interactions, ${\cal L}_{\rm int} = \theta y_f S \bar f f$, where $f$ is a charged fermion and $y_f$ is a Yukawa coupling. For $m_S > 2m_\mu$, the lifetime is 
\be
\tau_S \approx \frac{8\pi v^2}{\theta^2 m_\mu^2 m_S} \approx 10^{-2}\, {\rm s} \brac{10^{-7}}{\theta}^2 
\! \brac{\rm 500 \, MeV}{m_S},~~
\ee
which is a viable point in parameter space for a Higgs-mixed scalar \cite{Krnjaic:2015mbs,Dev:2020eam}.
Thus, the $S$ particle decay can be prompt relative to the timescale for BBN.

The parameters $\mu_S, A, \lambda_S$ and $V_0$ can be chosen such that $S = 0$ is a metastable vacuum, where $A^2  > 9 \lambda_S \mu_S^2/2$ ensures that 
 the vacuum at $S = v_S$ is the global minimum. 
 As a concrete example, for parameters 
\be
\label{eq:params}
A = 25.8 {\rm \, MeV} ~~,~~ \mu_S  = 10 {\rm \, MeV}~~,~~ \lambda_S = 1~~,
\ee
  the true vacuum satisfies  
 \be
v_S \approx \frac{1}{2\lambda_{S}} \left( A + \sqrt{A^2 - 4 \lambda_{S} \mu_S^2} \right) \approx 21 \, \rm MeV,
 \ee
 and the energy difference between vacua is 
\be
\Delta V =  \frac{\lambda_S}{4} v_s^4 - \frac{A}{6} v_S^3 \approx 9.6 \times 10^3 \, \rm MeV^4.
\ee
During EDE domination, the field is trapped at $S = 0$ but eventually tunnels to $S = v_S$ and reheats the universe through prompt $S \to \bar f f$ decays. In the instantaneous reheating approximation, we obtain
 \be
 \label{eq:TRH$A}
T_{\rm RH} = \brac{30 \, \Delta V}{\pi^2 g_\star}^{1/4}
\approx 10 \, {\rm MeV}\! \brac{\Delta V^{1/4}}{10 \, \rm MeV},
 \ee
 where we have used $g_\star = 10.75$ consistent with this temperature.

To determine the timescale for bubble nucleation, the bounce action $S_4$ can be approximated as \cite{Adams:1993zs}
\be
S_4 = \frac{4\pi^2
 (\alpha_1 \varepsilon + \alpha_2 \varepsilon^2 +\alpha_3 \varepsilon^3)
}{3\lambda_S (2- \varepsilon)^{3} }~~,~~ \varepsilon \equiv  \frac{9\lambda_S\mu_S^2}{A^2}~,~~~
\ee
where $\alpha_1 = 13.832$, $\alpha_2 = - 10.819$, and $\alpha_3 = 2.0765$. Using the parameters in Eq.~\eqref{eq:params}, $S_4 \approx 194$, so we estimate the timescale for bubble nucleation as
\be
\tau_{\rm PT} \sim \frac{H^3}{A^4} e^{S_4} \approx 0.3 \rm \, s~,
\ee
where we have used the Friedmann equation $3 H^2 = 8\pi G \Delta V$ in EDE domination. Although this timescale is naively quite close to the nominal BBN time, this occurs during EDE domination which alters the usual time-temperature relation.  From Eq.~\eqref{eq:TRH$A}, this choice of parameters the phase transition injects sufficient entropy to reheat the universe well above BBN temperature. Note that, because the phase transition timescale is exponentially sensitive to the parameters of the potential, it is possible to find a broad range of values compatible with the number of $e$-folds required to restore the observed baryon asymmetry before BBN.

Since the tunneling yields a first-order phase transition, gravitational waves will be generated from bubble collisions. For horizon sizes that correspond to $T_{\rm RH} \sim 10$ MeV, the present-day peak frequency of gravitational waves is $f_0 \sim$ nHz \cite{Kamionkowski:1993fg}, which is within the frequency range of pulsar timing array searches, though a detailed study of this signal is beyond the scope of this work.

\section{Baryon Yield Scaling}
\label{app:yield-calc}
Here we derive  final baryon yield $Y_B(a_{\rm RH})$ at reheating from Eq.~\eqref{eq:YB_RH}. justifying the nearly-horizontal contours in Fig.~\ref{fig:param-space}. 
If the Universe enters a baryon-dominated phase before the EDE domination (as shown in Fig.\ref{fig:timeline}), the energy density of the EDE and baryon components 
at $a_{\rm EDE}$ satisfies
\be
\label{eq:rhoBrhoEDE}
 \rho_B(a_{\rm EDE})=\rho_{\rm EDE},
\ee
where we have neglected the subdominant contribution from radiation and define $\rho_{\rm EDE} \equiv \rho_{\rm EDE}(a_{\rm EDE})$ without loss of generality. 
Assuming EDE domination ends instantaneously at $a = a_{\rm RH}$ when the EDE fluid is converted to radiation,  so we have 
\be
\label{eq:rho_EDE_RH}
\rho_{\rm EDE} \approx \rho_R(a_{\rm RH})  = \frac{\pi^2 g_\star( T_{\rm RH}) }{30}  T_{\rm RH}^4.
\ee
Since we can compute the baryon density at $a = a_{\rm EDE}$ using Eq. \eqref{eq:rhoBrhoEDE}, the baryon density after reheating is now determined by the number of e-folds, $N_e$,
in between $a_{\rm EDE}$ and $a_{\rm RH}$, 
\be
\label{eq:rhoB_RH}
 \rho_B(a_{\rm RH})=  \rho_B(a_{\rm EDE})\brac{a_{\rm EDE}}{a_{\rm RH}}^3   \approx \rho_{\rm EDE} \, e^{-3N_e} , ~~ 
\ee
where we have assumed full EDE domination in calculating the scale factor ratio in between $a_{\rm EDE}$ and $a_{\rm RH}$, thereby neglecting 
subdominant contributions from baryons and radiation during this epoch. 
Using the entropy density at reheating
\be
\label{eq:sRH}
s(a_{\rm RH})=  \frac{2\pi^2 g_{\star, s}(T_{\rm RH})  }{45} T_{\rm RH}^3~,
\ee
where $g_{\star,s} (T_{\rm RH}) = g_\star(T_{\rm RH})$ is the effective number of relativistic species for $T_{\rm RH} > $ MeV,
the final baryon-to-entropy ratio is 
\be
\label{eq:finalYB}
    Y_B(a_{\rm RH}) = 
 \frac{\rho_B(a_{\rm RH})}{s( a_{\rm RH}) }  
  \approx \frac{3 T_{\rm RH}}{2 m_p} e^{-3N_e}~,~
\ee
where we have used $n_B = \rho_B/m_p$ and Eqs. \eqref{eq:rho_EDE_RH}, \eqref{eq:rhoB_RH}, and \eqref{eq:sRH} to recover Eq.~\eqref{eq:YB_RH}

Thus, the final baryon asymmetry depends solely on $T_{\rm RH}$  and $N_e$ independently of the initial baryon asymmetry,  provided that the pre-existing radiation can be neglected during EDE domination. 
This property explains the horizontal contours in Fig.~\ref{fig:param-space} at large $Y_B(a_{\rm BG})$, during which the radiation can be safely ignored during this epoch. For smaller values of $Y_B(a_{\rm BG})$, the scale factor values $a_{\rm EDE}$ and $a_{\rm BD}$ are relatively close together, so the radiation component does not have sufficient time to redshift away, and therefore can not be neglected in this regime. This behavior introduces additional features for the contours of the e-fold number that yield the observed baryon asymmetry. We also note that, at large baryon chemical potential, the QCD phase diagram implies the hadronic confinement occurs at higher temperatures than the nominal $\Lambda_{\rm QCD} \approx 200$ MeV \cite{Kronfeld:2012uk}. However, in the large $Y_B({a_{\rm BG}})$ limit, the period of baryon domination is relatively long, so $a_{\rm EDE} \gg a_{\rm BD}$, and the approximations used in Eq.~\eqref{eq:finalYB} is valid.

\section{PNS Formation Criteria}
\label{app:formation}
In this appendix we adapt the formalism of PBH formation to our scenario in which a similar sequence of events transpires in a baryon dominated universe with the QCD equation of state. We begin by reviewing PBH formation with a constant sound speed  and then generalize the argument to a baryon dominated universe in which the sound speed varies over the course of evolution.

\subsection*{\bf Review of PBH Formation}
Here we review Sec 4b of Ref. \cite{Harada:2013epa} to estimate the perturbation threshold for PBH formation at horizon crossing. We consider an overdense region characterized by a perturbation $\delta = (\rho-\bar \rho)/\bar\rho$, where $\rho$ is the interior density and $\bar \rho$ is the background density. Inside this region, the metric satisfies
\be
ds^2 = dt^2 - \lambda^2 a(t)^2( d\chi^2 + \sin^2\!\chi \, d\Omega^2),
\ee
where $\chi$ is the comoving radius and $\lambda$ sets the curvature. The Friedmann equation in the overdensity is 
\be
\label{eq:fried}
H^2 = \frac{8\pi G \rho}{3} - \frac{1}{\lambda^2 a^2},
\ee
where $\rho$ is the density inside the overdense region. The background in which this region is embedded evolves according to
\be
\bar H^2 = \brac{\dot {\bar a}}{\bar a}^2= \frac{8\pi G \bar \rho}{3},
\ee
where $\bar H$ and  $\bar a$ are the background Hubble rate and scale factor, respectively.  We are interested in estimating the minimum perturbation size (at horizon crossing) for which a PBH will eventually form.

At horizon crossing, an overdensity with radius 
\be
\label{eq:Ra}
R(t) = \lambda a(t) \sin \chi~,
\ee
begins to expand and its interior density evolves as 
\be
\label{eq:rho_delta}
\rho = (1+\delta) \bar \rho,
\ee
where $\bar \rho \propto \bar a^{-3}$ and, assuming matter domination, $\delta \propto \bar a$. At the initial time $t = t_0$, we assume $\delta(t_0) < 1$, so  $\rho$ will decrease until the perturbation becomes nonlinear.  
The Jeans radius can be defined as   
\be
R_J = c_s \sqrt{ \frac{3}{8\pi G\rho} }~~,~~ c_s^2 =  \brac{dP}{d\rho}_s,
\ee
where $c_s$ is the sound speed, defined as the derivative at constant entropy. Initially, just after $t = t_0$, both $R$ and $R_J$ begin to grow. However, since the overdense region has positive curvature, the radial growth will eventually reach a maximum value at some later time $t = t_*$ where the interior scale factor satisfies $\dot a = 0$, so
\be
H^2 = 0 \implies \frac{8\pi G \rho_*}{3} = \frac{1}{\lambda^2 a_*^2},
\ee
where $a_* = a(t_*)$, $\rho_* = \rho(t_*)$ and, at this time, the Jeans scale becomes 
\be
\label{eq:RJstar}
R_{J*} = c_s \lambda a_*.
\ee
Initially, the Jeans scale grows faster than the radius of the overdensity, so if the Jeans criteria is satisfied at maximum radius, it's satified at all times. Therefore, the condition for collapse is $R_{*} > R_{J*}$. Using Eqs~\eqref{eq:Ra} and \eqref{eq:RJstar} we obtain
\be
c_s \lambda a_* < \lambda a_* \sin \chi~~ \implies ~~ \sin^2 \chi > c^2_s~,~ 
\label{eq:CarrJeansCriteria}
 \ee
and we can use Eq.\eqref{eq:Ra} to write
 \be
 \label{eq:sinXa}
\sin^2 \chi = \brac{R}{\lambda a}^2,
 \ee
 which is a purely geometric relation valid at all times. 
 We now wish to express this in terms of the initial $\delta(t_0)$ at horizon crossing.

 To this end, we define the time-dependent density fraction
$\Omega = 8\pi G \rho/(3H^2)$ and
  divide Eq.\eqref{eq:fried} by $H^2$ to find
 \be
\Omega  - 1 = \frac{1}{\lambda a^2 H^2} =  \frac{R^2_H}{\lambda a^2}  ,
 \ee
where we have used the definition of the horizon $R_H = H^{-1}$. Substituting this expression into Eq.~\eqref{eq:Ra} gives
 \be
 \label{eq:sinXa}
\sin^2 \! \chi = (\Omega -1) \brac{R}{R_H}^2,
 \ee
 and, to simplify further, we can use uniform Hubble slicing in which all regions have the same Hubble rate $H = \bar H$, so inside the overdensity we can write
 \be
\Omega = \frac{8\pi G\rho}{3\bar H^2 } = \frac{\rho}{\bar \rho},
 \ee
 where we have used $\bar H^2 = 8\pi G\bar \rho/3$ in the last step. From the definition of $\delta$, we have 
 \be
\delta = \frac{\rho - \bar \rho}{\bar \rho}  \implies \Omega =  \frac{\rho}{\bar \rho} = 1+\delta ~,
 \ee
 so Eq.~\eqref{eq:sinXa} can be written  
 \be
 \label{eq:sin2}
\sin^2 \! \chi = \delta(t) \brac{R(t)}{R_H(t)}^2,
 \ee
 where we have restored explicit time dependence.
 Using the fact that $\sin \chi$ is a purely geometric quantity, which is valid at all times,  we can evaluate this expression at horizon crossing where $R(t_0) = R_H(t_0)$, so we combine  Eqs. \eqref{eq:sin2} and \eqref{eq:CarrJeansCriteria} to eliminate $\sin^2\chi$ and obtain
 \be
 \label{eq:jeans_final}
 \delta_H \equiv \delta(t_0) > c_s^2 ,
 \ee
which identifies the critical {\it initial} overdensity at horizon crossing required to eventually collapse to form a PBH.

Note that argument here is based on Sec. 4b of Ref. \cite{Harada:2013epa}, which 
reviews the original PBH formation argument from Carr in Ref. \cite{Carr:1974nx}. Although Ref. \cite{Harada:2013epa}, 
updates and improves this  this earlier analysis to obtain 
a more rigorous collapse, threshold, the end result is qualitatively similar and only differs at the ${\cal O}(10\%)$ from the Carr analysis, so we appropriate the simpler earlier analysis for our purposes.

\subsection*{PNS Formation}
We now adapt this argument to estimate the criteria for PNS formation in a baryon dominated universe, so we identify $\rho \approx \rho_B$, neglecting other subdominant components. As above, we assume that a spherically uniform overdensity enters the horizon at $t=t_0$ with initial radius $R(t_0) = \lambda a(t_0) \sin \chi$.
However, unlike in the preceding PBH discussion, here the equation of state depends on both temperature and density, so the sound speed is not constant and we must account for its evolution.

 During baryon domination, the horizon mass satisfies 
 \be
 M_H =  \brac{3}{32 \pi G^3 \rho_B}^{1/2}   \approx 1M_\odot \brac{10^2 \rho_{\rm nuc}}{\rho_B}^{1/2} \!\!\!,~~~
 \ee
 so the universe must exceed nuclear density at $t = t_0$ when a candidate perturbation $\delta$ enters the horizon, otherwise the enclosed mass will exceed the Tolman-Oppenheimer-Volkoff limit of $\sim 2 M_\odot$  \cite{Tolman1939,OppenheimerVolkoff1939} when it eventually collapses and forms a PBH instead. Throughout our analysis, we are interested in the $M_H \in (0.1-2) M_\odot$ range, compatible with neutron star stability \cite{Douchin:2001sv,Bordbar:2006ij}.

 As in the PBH case discussed above, at $t = t_* > t_0$, the  perturbation now expands to its maximum radius $R(t_*) = R_*$  and the density becomes
 \be
\rho_B(t_*) \approx \rho_B(t_0) \brac{a_0}{a_*}^3 \ll \rho_{\rm nuc},
 \ee
 where $a_i = a(t_i)$ and we have assumed that $\delta(
t_*) < 1$ in order to neglect the additional $(1+\delta)$ factor from Eq.~\eqref{eq:rho_delta}. For sub-nuclear densities at $t = t_*$, we adopt an approximate sound speed dominated by thermal and electron-degeneracy pressure,\footnote{The formula for electron degeneracy pressure is evaluated at lepton fraction $n_e/n_B = 1/2$.} which satisfies
\be
\label{eq:cs-app}
c_{s}^2(t_*) \approx \frac{5T_*}{3m_p} + \frac{4\rho_{R*}}{9\rho_{B*}}  + \frac{1}{3} \left( \frac{ 3\pi^2 \rho_{B*} }{ 16 m_p^4}  \right)^{1/3}\!\!~.
\ee
 From Eq.~\eqref{eq:jeans_final}, the Jean criterion is $\delta_H > c_{s}^2(t_*)$, so  any perturbation that satisfies this 
inequality will at least begin to undergo inward contraction. However, as noted in the main text, during collapse the nuclear equation of state stiffens, resulting in $c^2_{s, \rm nuc} \sim 0.2-0.6$ \cite{Fujimoto:2024cyv,Fukushima:2025ujk}. Thus,
for perturbations in the range 
\be
\label{eq:jeans-app}
c^2_{s,*} \lesssim \delta_H \lesssim c_{s,\rm nuc}^2~,
\ee
this initial collapse phase encounters a pressure response and may settle into a stable compact object.

At horizon crossing $R = R_H$, so Eq.~\eqref{eq:sin2} yields 
$\sin\chi = \sqrt{\delta_H}$, and Eq.~\eqref{eq:Ra} beomces
\be
R = \lambda \sin \chi = H_0^{-1}~~, ~~   \lambda = \frac{1}{\sqrt{\delta_H} H_0} ~, 
\ee
where we have taken $a_0 =1$, and $H_0 \equiv H(t_0)$, not to be confused with the present-day Hubble constant. The Friedmann equation in Eq.~\eqref{eq:fried} can now be written 
\be
H^2 = \frac{8\pi G \rho_B}{ 3} - \frac{\delta_H H_0^2}{a^2}
\ee
and at the maximum radius $H = 0$, so we have 
\be
\label{eq:fried-max}
\frac{8\pi G \rho_{B*}}{ 3} =
 \frac{\delta_H H_0^2}{a_*^2}.
\ee
Adopting constant Hubble slicing $H = \bar H$, and assuming $c_s \ll 1$, we can write 
\be
\rho_{B*} = \frac{\rho_{B,0}}{ a_*^{3}} = \frac{\bar \rho_{B,0}}{a_*^3} (1+\delta_H) =\frac{3\bar H_0^2}{8\pi G a_*^3} 
(1+\delta_H),~~~~~~
\ee
where $\bar\rho_{B,0} = \bar \rho_B(t_0)$ is the unperturbed background density. Using this result, 
Eq.~\eqref{eq:fried-max} 
yields
\be
\label{eq:astar}
a_* = \frac{1+ \delta_H}{\delta_H},
\ee
which is exact in the limit that the interior region redshifts as non-relativistic matter. This condition will only be approximately true in our scenario since the sound speed is nonzero throughout the overdensity evolution, but is typically small near the maximum radius. 
The result in Eq~\eqref{eq:astar} allows us to write $\rho_{B*}$ and $\rho_{R*}$ as functions of $\delta_H$ in Eq.~\eqref{eq:cs-app} to evaluate the Jeans criterion in Eq.~\eqref{eq:jeans-app}.

\begin{figure}
    \centering
    \includegraphics[width=0.95\linewidth]{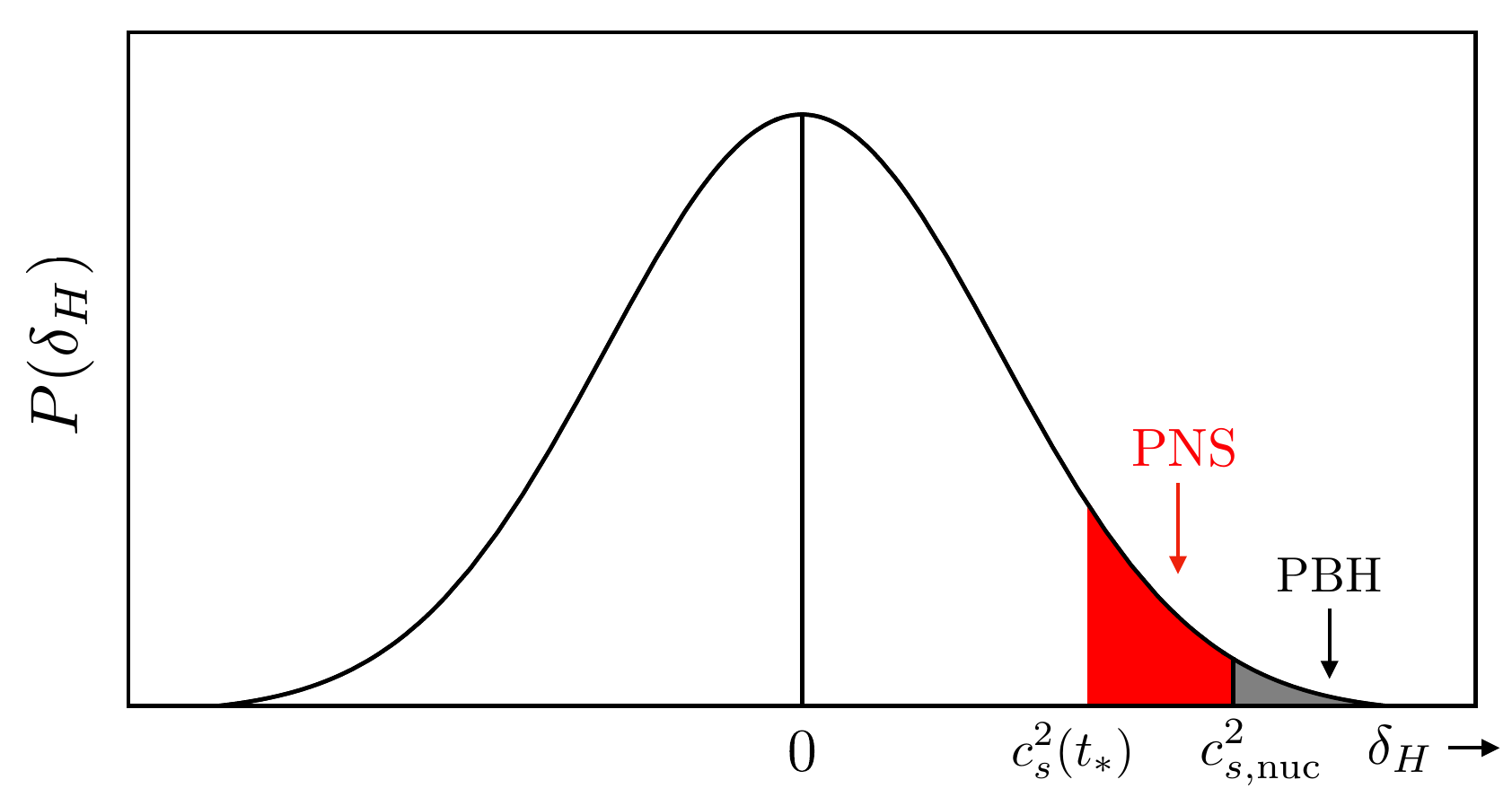}
    \caption{Schematic probability distribution of density fluctuations $\delta_H$ at horizon crossing during  baryon domination when $M_H \sim M_\odot$. The red window corresponds to the range from Eq.~\eqref{eq:jeans-app}, in which PNS formation may occur, and the gray region with $\delta_H > c^2_{s, \rm nuc}$ corresponds to the PBH formation range.be}
    \label{fig:pdf}
\end{figure}

\subsection*{Abundance Estimate}

For PBH formation in standard cosmology, the fractional PNS abundance is governed by $P(\delta_H)$, the probability distribution of primordial functions. As noted in   Eq.~\eqref{eq:jeans_final}, any region with  super-critical values of $\delta_H > c_s^2$ results in a PBH. Thus, assuming Gaussian perturbations at horizon crossing, the PBH abundance at the time of formation is  \cite{Green:2020jor} 
\be
\label{eq:int-pbh}
f_{\rm {PBH} }  = \frac{\rho_{\rm PBH}}{\rho_{\rm tot}} 
= \int_{\delta_{H,c}}^\infty 
d\delta_H P(\delta_H),
\ee
where $\rho_{\rm tot} = 3 H^2/(8\pi G)$ is the total energy density, $\delta_{H, c} \approx c_s^2$ is the critical overdensity for PBH formation for a constant sound speed, and 
\be
\label{eq:prob}
P(\delta_H) = \frac{1}{\sqrt{2\pi \sigma^2}} \exp{\left(-\frac{ \delta_H^2}{2\sigma_H^2 } \right)},
\ee
where $\sigma_H$ is the variance of perturbations on scale $R_H$, which is set by the primordial power spectrum \cite{Blais:2002gw,Josan:2009qn}. 
For PNS formation, the abundance fraction is 
\be
\label{eq:int-pns}
f_{\rm {PNS} }  = \frac{\rho_{\rm PBH}}{\rho_{B}} 
= \int_{c^2_{s}(t_*)}^{c^2_{s,\rm nuc}} 
d\delta_H P(\delta_H),\\ \nonumber
\ee
where $P(\delta_H)$ is the same probability distribution from Eq.~\eqref{eq:prob}, but here we integrate only over the finite range in
Eq.~\eqref{eq:jeans-app} and identify $\rho_{\rm tot} = \rho_B$ since PNS formation can only occur during baryon domination. 

In Fig. \ref{fig:pdf}
 we show a schematic plot of the PNS formation range. Note that the integrals in Eqs.~\eqref{eq:int-pbh} and \eqref{eq:int-pns} are exponentially sensitive to the sound speeds $(c_{s,*}, c_{s,\rm nuc})$ and the primordial power spectrum, which respectively govern the integration limits and the distribution variance. Thus, only a dedicated simulation can reliably predict the relative PNS/PBH abundances and we leave this to future work.

\bibliographystyle{utphys3}
\bibliography{biblio}

\end{document}